\documentclass[letterpaper]{article}

\usepackage[T1]{fontenc}

\usepackage{geometry}
\usepackage{setspace}
\usepackage{subcaption}
\usepackage{booktabs}

\usepackage{achemso}

\usepackage{graphicx}
\usepackage{float}
\newfloat{scheme}{htbp}{los}
\floatname{scheme}{Scheme}
\floatname{chart}{Chart}
\newfloat{graph}{htbp}{loh}

\usepackage{chemformula} % Formulas using \ch{}
\usepackage[version = 4]{mhchem} % Formulas using \ce{}

\usepackage{authblk}
\author[1]{Nathan Sitaraman*}
\author[2]{Zhaslan Baraissov}
\author[1]{Alexis Grassl}
\author[2]{Hongbin Yang}
\author[2]{Daniel Tong}
\author[2]{David A. Muller}
\author[1]{Matthias Liepe}
\affil[1]{Cornell Laboratory for Accelerator-based ScienceS and Education, Ithaca, USA}
\affil[2]{School of Applied Physics, Cornell, Ithaca, USA}

\title{Synthesis and Characterization of Atomically-Sharp Superconductor-Dielectric Interface}
\date{*Email: nathan.sitaraman@gmail.com}

\begin{document}

\maketitle

\begin{abstract}
  Modification of superconductor-dielectric interfaces is known to strongly impact coherence times of superconducting quantum devices. This relationship is thought to arise from differences in the concentration of "two-level system" defects in the disordered dielectrics and superconductor-dielectric interfaces; these defects couple to electromagnetic modes in the device and cause dissipation. Zirconium oxide barrier layers on niobium have emerged as a promising pathway to low-loss interfaces in recent years, evidently due to  the crystalline nature of these layers in comparison to the amorphous niobium native oxide. We explain the unique ability of zirconium oxide to form a crystalline layer, to maintain a sharp interface with metallic niobium, and to prevent niobium oxide re-growth in terms of the chemical properties of ZrO$_2$ and the Nb-Zr-O ternary system. We demonstrate a new method to grow air-stable zirconium oxide layers on niobium with a higher level of crystallinity and a sharper oxide-metal interface than previously shown, and provide the first comprehensive microscopic analysis of ZrO$_2$ capping layer properties. These developments pave the way toward vital performance advances in superconducting quantum devices.
\end{abstract}

\section*{Keywords}

Superconducting quantum electronics, niobium superconducting resonators, metal-oxide interfaces, thin oxide characterization, x-ray photoelectron spectroscopy, multislice electron ptychography

\section{Introduction}

Superconducting radio-frequency (SRF) resonators have historically been developed for applications in particle accelerators, but in recent years there has been increasing interest in developing SRF technology for quantum applications~\cite{hollister2022large,romanenko2020three,PRXQuantum.4.030336,bal2024systematic,siddiqi2021engineering,blais2021circuit}. Unlike SRF cavities used in particle accelerators, resonators for quantum applications operate at milliKelvin temperatures and very low fields; optimizing performance in this fundamentally different regime of SRF physics comes with new challenges and demands new solutions~\cite{heidler2021non,kalboussi2024reducing,kalboussi2025crystallinity}.

Under the low-field, milliKelvin temperature conditions relevant for quantum computing, RF losses are dominated by the so-called “residual” resistance, as opposed to BCS losses arising from thermally-excited quasiparticles in the superconductor. The residual resistance arises from non-superconducting phases that exist at interfaces such as the superconductor surface; these phases can interact with the RF field and cause dissipation.

The dissipation arising from non-superconducting surface phases at very low fields has often been described with a two-level system (TLS) model~\cite{muller2019towards,o2008microwave,romanenko2017understanding,martinis2005decoherence}. Specifically, we may consider a localized electronic state which is very close to the Fermi energy but does not participate in the condensate. If this state can be driven to an excited state by absorbing an amount of energy similar to the photon energy of the resonator, then this constitutes a TLS loss mechanism. More generally, TLS excitations may involve some combination of electronic and nuclear degrees of freedom; the only constraint is that a process exists for them to absorb RF photons.

The primary difficulty in eliminating TLS loss mechanisms from a superconducting device is the presence of amorphous “native” oxides. All elemental superconductors used for superconducting resonators form thin, disordered native oxides within seconds, even under high-vacuum conditions. In this paper we will focus on niobium, which along with aluminum has been the most widely-used superconductor for RF applications~\cite{blais2021circuit}.

\begin{figure}[ht]
\centering
\begin{subfigure}{.5\textwidth}
\centering
\includegraphics[width=0.95\textwidth]{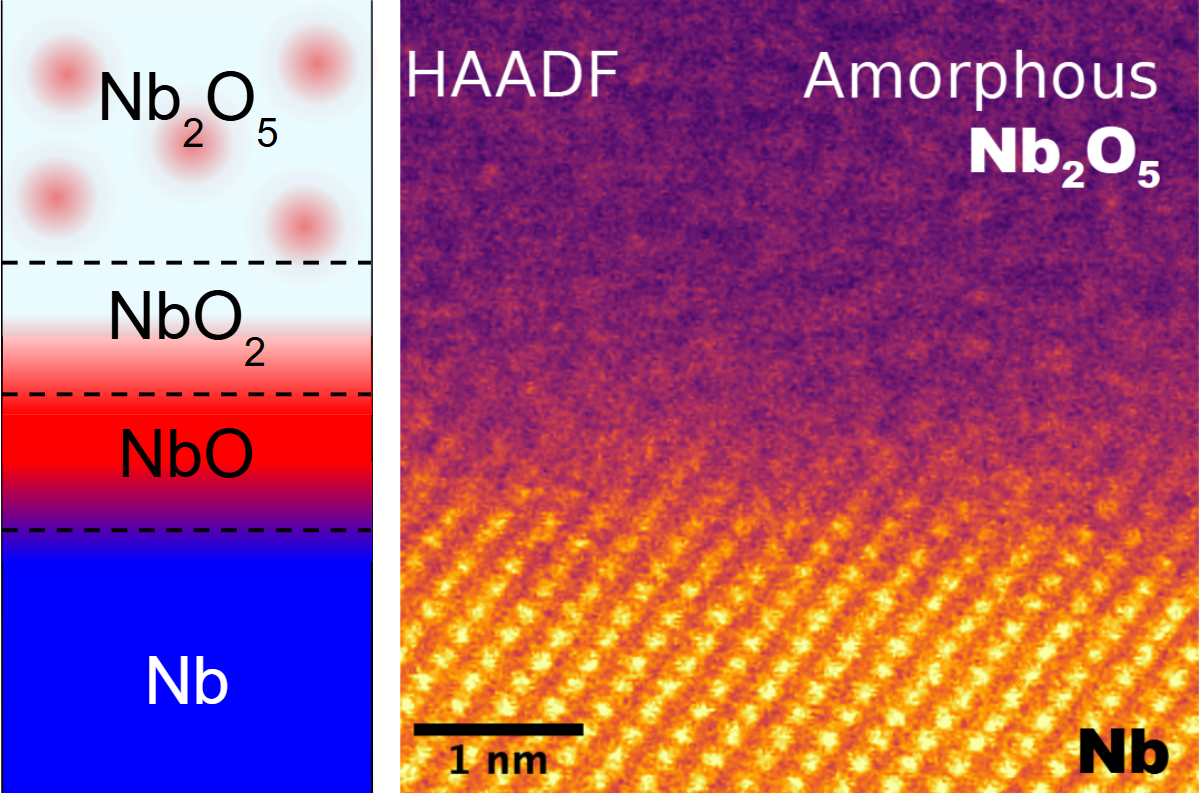}
\caption{NbO$_x$/Nb interface}
\label{fig1:sub1}
\end{subfigure}%
\begin{subfigure}{.5\textwidth}
\centering
\includegraphics[width=0.95\textwidth]{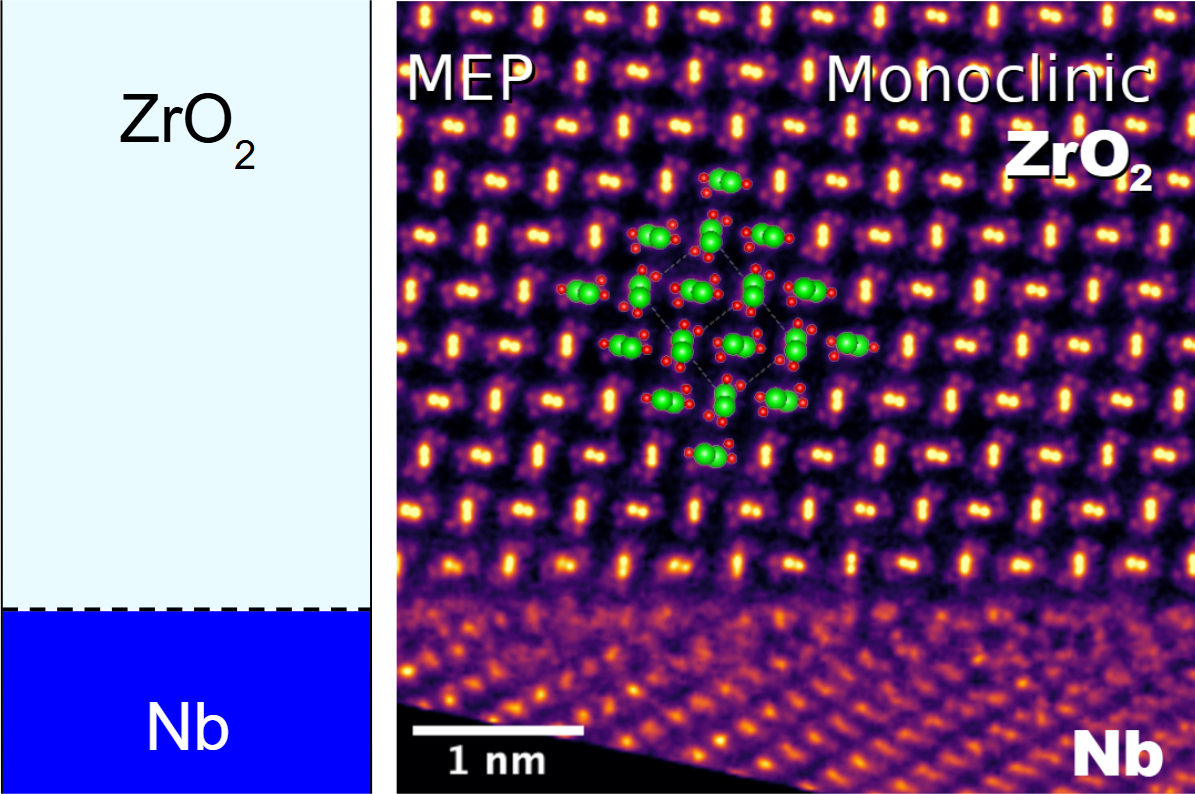}
\caption{ZrO$_2$/Nb interface}
\label{fig1:sub2}
\end{subfigure}
\caption{Comparison between (a) the NbO$_x$/Nb interface and (b) the ZrO$_2$/Nb interface. The schematics show expected distributions of normal-conducting electrons (red) and Cooper pairs (blue). Annular dark-field (ADF) cross section shows a gradual interface between Nb and amorphous Nb$_2$O$_5$, while multislice electron ptychography  (MEP) shows a sharp interface between Nb and ZrO$_2$.}\label{fig1}
\end{figure}

The literature generally recognizes three main niobium oxide phases: NbO, NbO$_2$, and Nb$_2$O$_5$. The native niobium oxide contains all three of these phases in layers, with the fully-oxidized Nb$_2$O$_5$ on the surface, followed by NbO$_2$, and finally NbO at the interface with elemental niobium. Recent studies have helped build evidence that these oxide phases, and the Nb$_2$O$_5$ phase in particular, are responsible for TLS losses~\cite{kalboussi2025crystallinity,wenskat2022vacancy,bafia2024oxygen}.

While the exact microscopic mechanism responsible for TLS losses in Nb$_2$O$_5$ is not yet known, both theoretical and experimental studies suggest that defects in the oxide play a key role. The native oxide is amorphous, consisting primarily of Nb$_2$O$_5$ oxide, which is expected to have a very high defect density and a correspondingly rich array of possible excitations~\cite{kalboussi2025crystallinity}. Additionally, a variety of non-stoichiometric Nb$_2$O$_{5-\delta}$ phases have been described, which generally contain conduction-band electrons that could absorb RF photons~\cite{koccer2019first}. A hypothetical ideal oxide would in contrast be perfectly crystalline, with no conduction-band electrons (or valence band holes) and no possible excitations near the RF photon energy.

The understanding that a more crystalline oxide could improve device performance has led us to consider ZrO$_2$ as a capping layer for niobium, due to a constellation of useful properties that this manuscript will describe in full. In a few short years since ZrO$_2$ capping layers were first proposed, independently by researchers at Cornell and at Paris-Saclay, ZrO$_2$ has quickly become a material of choice for this application~\cite{sitaraman2022theory,kalboussi2023nano,sitaraman2023zr,choi2025low,kalboussi2023surface,kalboussi2025crystallinity2}. Here, we present a recipe that produces a far more crystalline ZrO$_2$ capping layer than has previously been demonstrated for any oxide on niobium, and an unusually-sharp superconductor-dielectric interface  (Fig.~\ref{fig1}). 

\section{Results}

\begin{figure}[ht]
\centering
\includegraphics[width=0.7\textwidth]{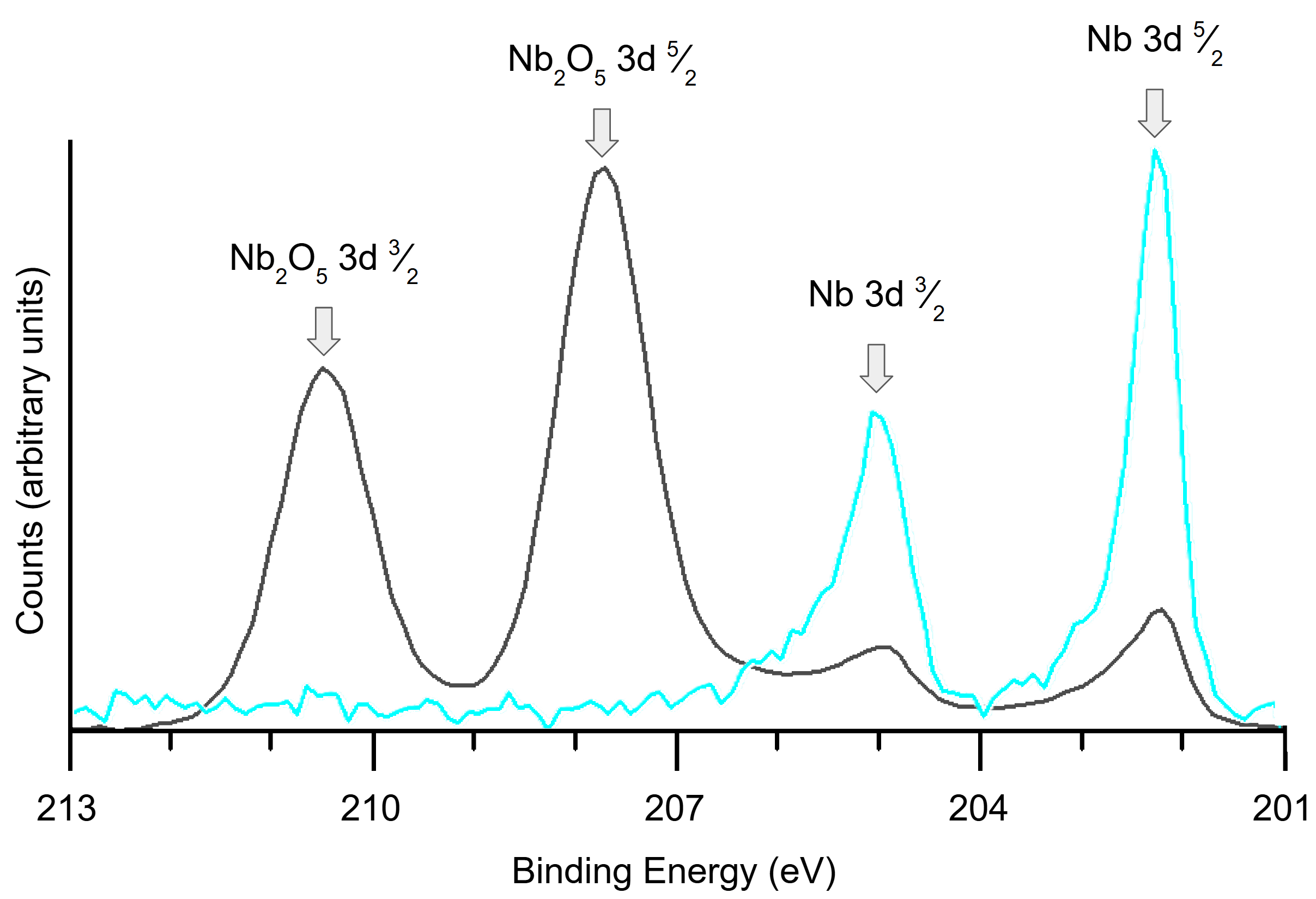}
\caption{Comparison of Nb XPS spectrum for an electropolished Nb reference sample (black) and the ZrO$_2$ capping layer processed at 800C (teal), with peak positions labeled for oxide and metallic peaks. In contrast to the reference sample, the ZrO$_2$ capping layer sample shows only metallic Nb peaks. Background features have been subtracted, and spectra have been rescaled for ease of comparison.}\label{fig2}
\end{figure}

Our study begins by using X-ray photoelectron spectroscopy (XPS) to explore the effects of some common vacuum baking recipes for niobium SRF cavities on a Zr-coated niobium surface, and to assess effectiveness of the ZrO$_x$ passivation layer at preventing re-oxidation of Nb in comparison to standard-preparation electropolishing (Fig.~\ref{fig2}). After developing an understanding of how the surface evolves at progressively higher temperatures and confirming surface passivation, we further investigate our most promising sample using scanning electron microscopy (SEM) and cross-section scanning transmission electron microscopy (STEM) to analyze the structure of the passivation layer and the superconductor-dielectric interface in greater detail.

\subsection{X-Ray Photoelectron Spectroscopy}\label{subsec1}

XPS was used to analyze four samples: an ``as-deposited" sample which was not subjected to vacuum annealing, a ``120C" sample which was baked according to the Fermilab 2-step process, an ``800C" sample which was baked for 5 hours at 800C, and an ``1100C" sample which was baked for 5 hours at 1100C~\cite{grassellino2018accelerating}. Samples were analyzed shortly after exposure to atmosphere at room temperature and then re-analyzed after many months. Based on results from our earlier study, we maintained a deposition thickness of 4nm metallic Zr across all samples~\cite{sitaraman2023zr}.

\begin{figure}[ht]
\centering
\begin{subfigure}{.45\textwidth}
\centering
\includegraphics[width=1.0\textwidth]{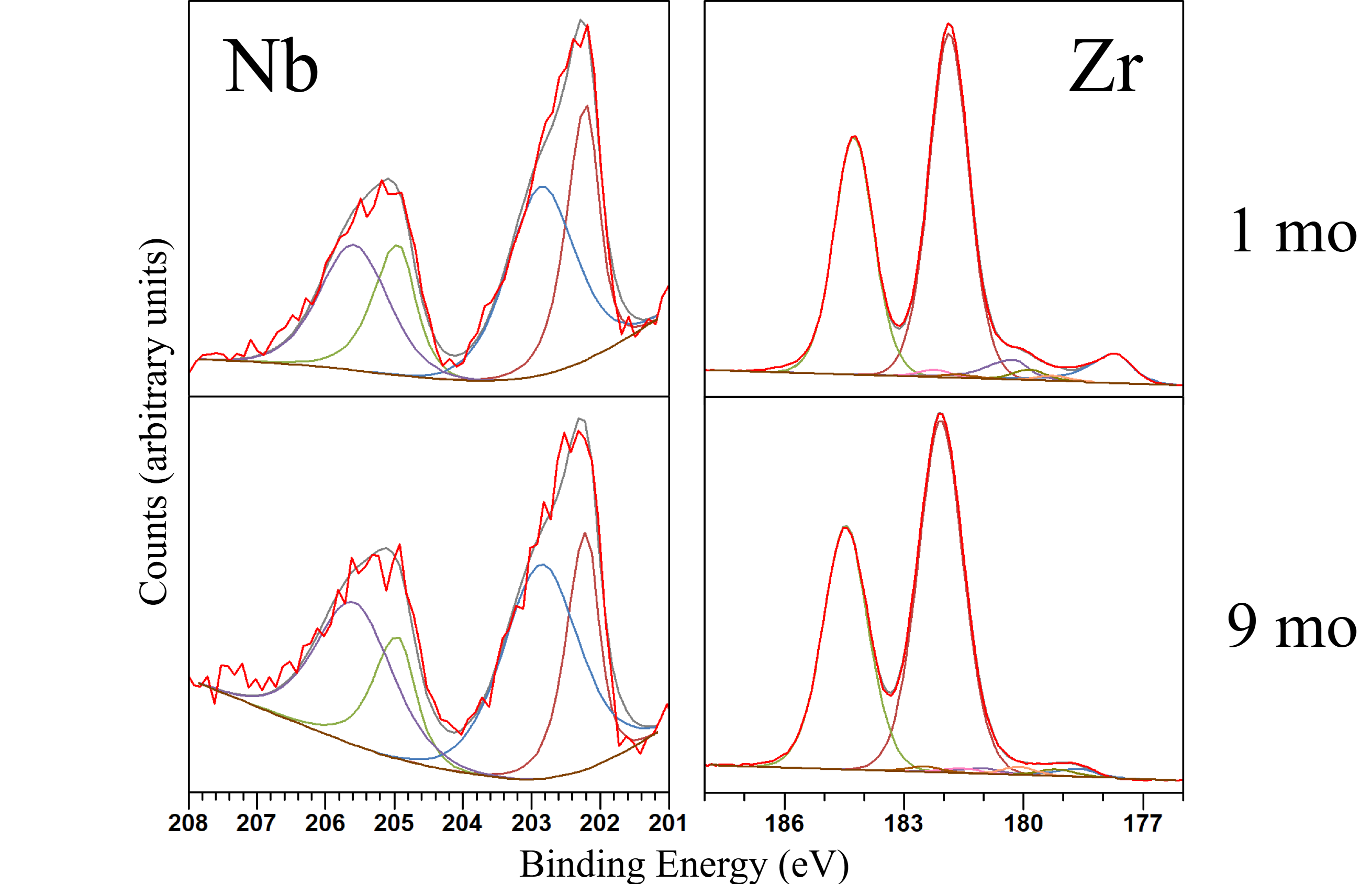}
\caption{As-deposited}
\label{fig3:sub1}
\end{subfigure}%
\begin{subfigure}{.45\textwidth}
\centering
\includegraphics[width=1.0\textwidth]{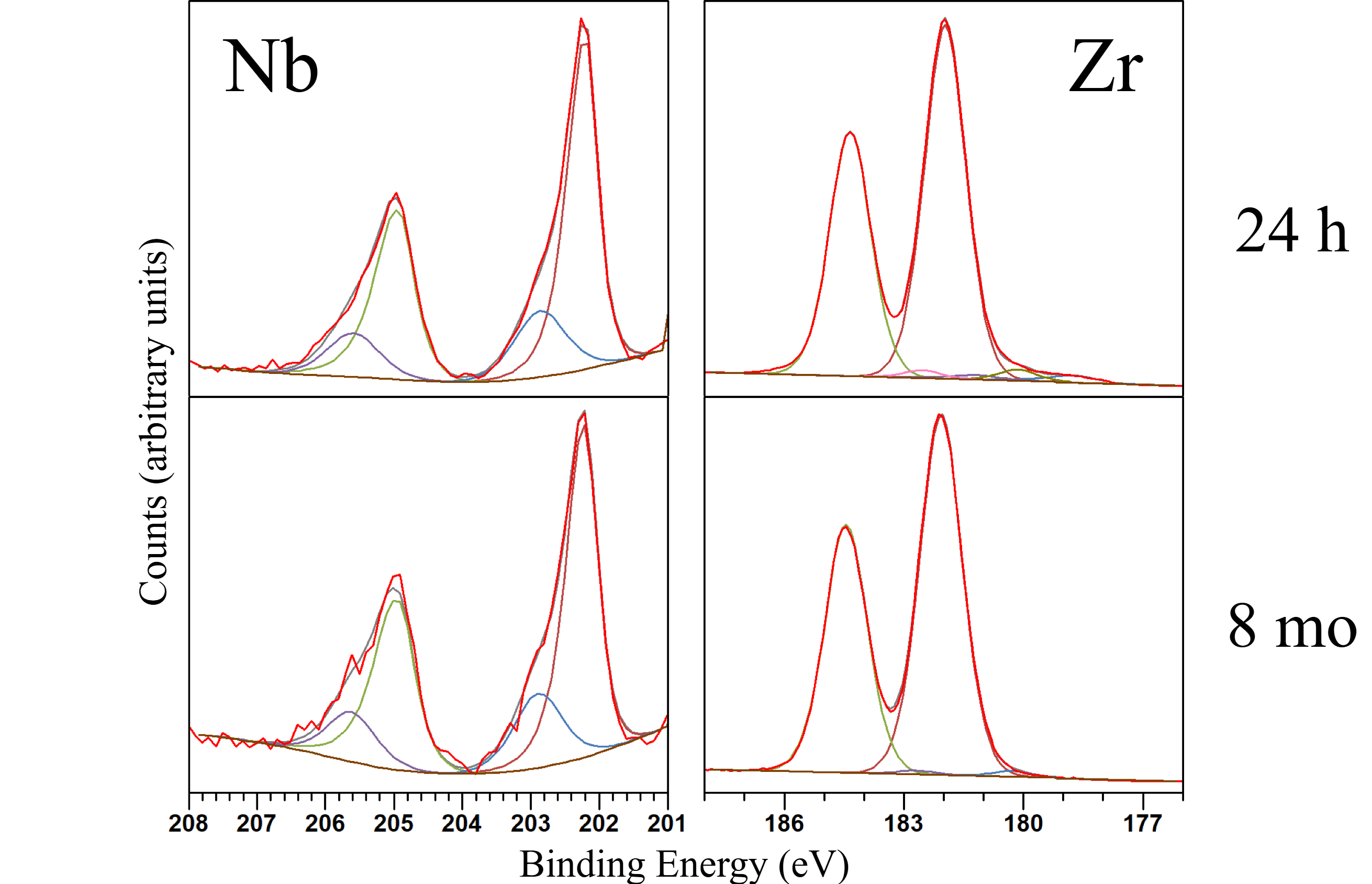}
\caption{120C-baked}
\label{fig3:sub2}
\end{subfigure}
\caption{Nb and Zr XPS spectra for as-deposited and 120C-baked samples, showing diminished suboxide components for both Nb and Zr after 120C baking.}\label{fig3}
\end{figure}

First considering as-deposited and 120C-baked samples, we see complete suppression of the Nb$_2$O$_5$ phase in the Nb elemental spectrum. The 3d $\frac{5}{2}$ Nb$_2$O$_5$ doublet peak would be expected at a binding energy between 207 and 208eV, but no peak is observed at this energy (Fig.~\ref{fig3})~\cite{dacca1998xps}. Instead, only a single doublet is observed in all cases, corresponding to overlapping suboxide and metal peaks. The broader suboxide peaks exist at a slightly higher binding energies than the narrower, asymmetric metallic peaks. The fitted peak areas are relatively unchanged after long exposure to atmosphere, indicating that atmospheric oxygen was not able to diffuse through the capping layer to react with the Nb substrate on these timescales.

Notably, the 120C-baked sample showed a relatively weaker Nb suboxide signal in comparison to the as-deposited sample (Fig.~\ref{fig3:sub2}). This suggests that short-range diffusion at 120C allowed lingering metallic Zr to react with residual oxygen in the Nb substrate, resulting in a more metallic Nb spectrum.

In the Zr elemental spectrum, we see the most intense peaks occur at a high binding energy corresponding to the high oxidation state of Zr atoms in the primary ZrO$_2$ oxide. This indicates that the majority of metallic Zr deposited on the surface has fully oxidized. Peaks of lower oxides occur at lower binding energies, and the unoxidized Zr metal peaks occur at the lowest binding energy of all; contributions from many such peaks, including prominent metallic peaks, are most apparent in the as-deposited sample analyzed after 1 month of exposure to atmosphere.

Our results show that both 120C baking and long exposure to atmosphere encourage the continued oxidation of the Zr layer. This could be explained either by the diffusion of gaseous oxygen through the primary oxide, or by the gettering of oxygen in the niobium substrate by the capping layer. Notably, the widths of the ZrO$_2$ peaks in the as-deposited sample increased by a small but significant amount after long exposure to atmosphere, possibly indicating the gradual emergence of different ZrO$_2$ phases or microstructures, resulting in slight differences in XPS binding energies for Zr atoms in subtly-different bonding configurations.

\begin{figure}[ht]
\centering
\begin{subfigure}{.45\textwidth}
\centering
\includegraphics[width=1\textwidth]{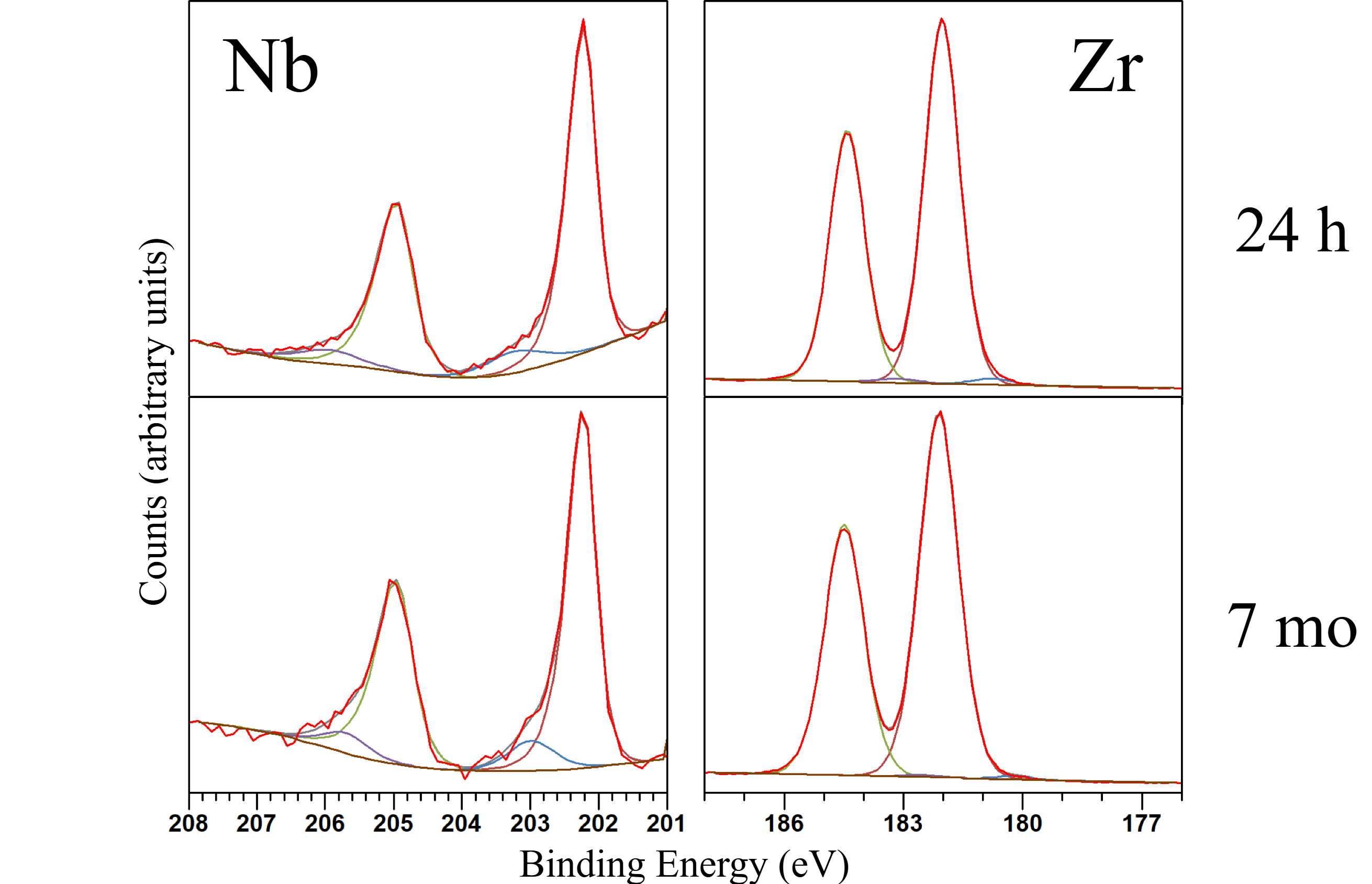}
\caption{800C-baked}
\label{fig4:sub1}
\end{subfigure}%
\begin{subfigure}{.45\textwidth}
\centering
\includegraphics[width=1\textwidth]{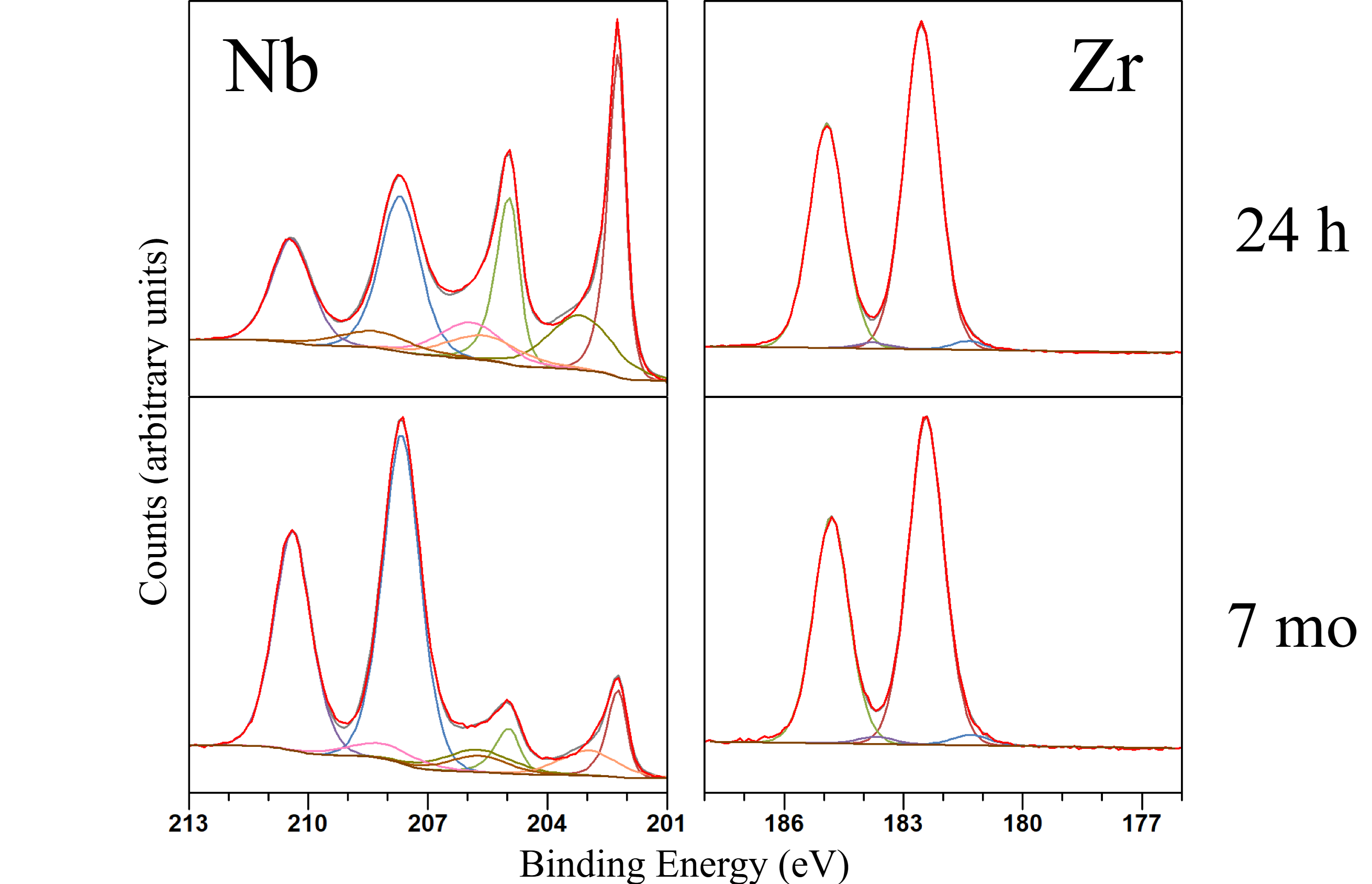}
\caption{1100C-baked}
\label{fig4:sub2}
\end{subfigure}
\caption{Nb and Zr XPS spectra for 800C-baked and 1100C-baked samples, showing greatly-diminished suboxide components for the 800C-baked sample, and showing the re-emergence of prominent Nb$_2$O$_5$ peaks for the 1100C-baked sample.}\label{fig4}
\end{figure}

Next considering the 800C-baked and 1100C-baked samples, we see that the 800C-baked sample shows only small shoulder peaks indicative of Nb suboxide on the tails of very clear, sharp Nb metal peaks  (Fig.~\ref{fig4:sub1}). The areas of such small peaks are difficult to ascertain, as they are quite sensitive to the choice of background (complicated in this case by the presence of Zr plasmon features) and to the choice of peak shape parameters. Our peak deconvolution attempts to be conservative in this regard, favoring a stronger suboxide peak, lower background, and weaker Nb metal asymmetry (tail). It is possible therefore that the actual suboxide component is even smaller than what is proposed here. After 7 months of exposure to atmosphere, the magnitude of this shoulder feature had subtly but clearly increased, suggesting a very slow but detectable diffusion of oxygen through the capping layer.

The 1100C-baked sample in contrast showed more significant changes in surface composition relative to samples processed at lower temperatures (Fig.~\ref{fig4:sub2}). A Nb$_2$O$_5$ component appeared, indicating that the capping layer is no longer effectively passivating the substrate. This could be caused by evaporation of ZrO$_2$, diffusion of Zr into the substrate, or the formation of pores in the capping layer due to lateral diffusion. After 7 months of exposure to atmosphere, the Nb$_2$O$_5$ component increased significantly in magnitude while the Nb metal component decreased significantly in magnitude, as expected for a niobium substrate unprotected from atmospheric oxygen.

In the Zr spectrum, the 800C sample shows a further reduction in the magnitude of the suboxide component relative to the 120C sample. This is as expected, considering that at this temperature any kinetic limitations preventing oxidation at the interface between the substrate and the capping layer should be overcome. Notably, after 7 months of exposure to atmosphere, this suboxide component was even further reduced, to the point of being almost undetectable. This observation hinted at the possibility that the suboxide layer may be completely absent at least at some locations on the Nb-ZrO$_2$ interface, and motivated the cross-section analysis of the interface described in the following subsection. This may also reflect the gradual filling of energetically-marginal oxygen sites at ZrO$_2$ grain boundaries.

The Zr spectrum of the 1100C-baked sample is unremarkable, aside from being lower-intensity than that of samples processed at lower temperatures. This, again, could be explained by a layer that is significantly thinner, or discontinuous, or both. The narrow width of the ZrO$_2$ peaks could indicate a higher degree of crystallinity in the remnants of the capping layer, or a different crystal phase.

In order to interpret our XPS results quantitatively, we choose to focus on a few figures of merit summarized in Table~\ref{tab1}. First, we consider the ratio of suboxide signal to Nb metal signal. These XPS signals are attenuated proportionately by the presence of the primary oxide, while the Nb metal signal is further attenuated by the presence of the suboxide layer. Therefore, the ratio of (Nb plus Zr) suboxide signal to Nb metal signal is a sensitive indicator for the total amount of suboxide phases that exist between the substrate and the primary oxide. Spectra from the 800C-baked sample exposed to atmosphere for 7 months show a significantly lower ratio of suboxide to Nb metal than spectra for any other sample.

\begin{table}[ht]
\caption{Quantitative analysis of XPS data}\label{tab1}%
\begin{tabular}{@{}llll@{}}
\toprule
Sample & (Suboxide)/(Nb Metal)  & (Nb Metal)/(Total Oxide) & ZrO$_2$ Peak Width (eV)\\
\midrule
As-Deposited (24h)\footnotemark[1]    & 4.53   & 0.010  & 1.17  \\
120C Baked (24h)\footnotemark[1]    & 1.22   & 0.031  & 1.22  \\
800C Baked (24h)    & 0.79   & 0.025  & 1.04  \\
1100C Baked (24h)\footnotemark[2]    & 0.88   & 0.403  & 1.02  \\
As-Deposited (9mo)\footnotemark[1]    & 5.43   & 0.009  & 1.32  \\
120C Baked (8mo)    & 0.87   & 0.026  & 1.25  \\
800C Baked (7mo)    & 0.39   & 0.025  & 1.16  \\
1100C Baked (7mo)\footnotemark[2]    & 1.74   & 0.091  & 1.04  \\
\bottomrule
\end{tabular}
\footnotetext[1]{Zr metal present in addition to suboxide.}
\footnotetext[2]{Nb$_2$O$_5$ present in addition to ZrO$_2$.}
\end{table}

Next, we consider the ratio of Nb metal signal to the total oxide signal. This ratio gives us information about the overall thickness and uniformity of the capping layer. A thicker and/or more uniform capping layer is expected to have a smaller ratio, and vice versa. Compared to the 120C-baked sample, the 800C-baked sample shows little change in ratio of Nb metal to total oxide, indicating that evaporation and dissolution reactions are still extremely slow at 800C.

Finally, we consider the ZrO$_2$ peak width. Variations in this width give us some insight into the structure of the ZrO$_2$ layer. Specifically, while different crystal structures of ZrO$_2$ are too similar in XPS binding energy to fully deconvolve, the presence of multiple phases could affect the width of the overall ZrO$_2$ peak. Alternatively, a broader ZrO$_2$ peak could be indicative of variations in the Fermi level of the substrate relative to the ZrO$_2$ conduction band edge at different locations in the sample, resulting in slightly different band-bending effects acting on the ZrO$_2$ peak position~\cite{lackner2019using}. With the exception of 1100C-baked samples, all recipes show some degree of ZrO$_2$ peak broadening.

\begin{figure}[ht]
\centering
\begin{subfigure}{.45\textwidth}
\centering
\includegraphics[width=0.9\textwidth]{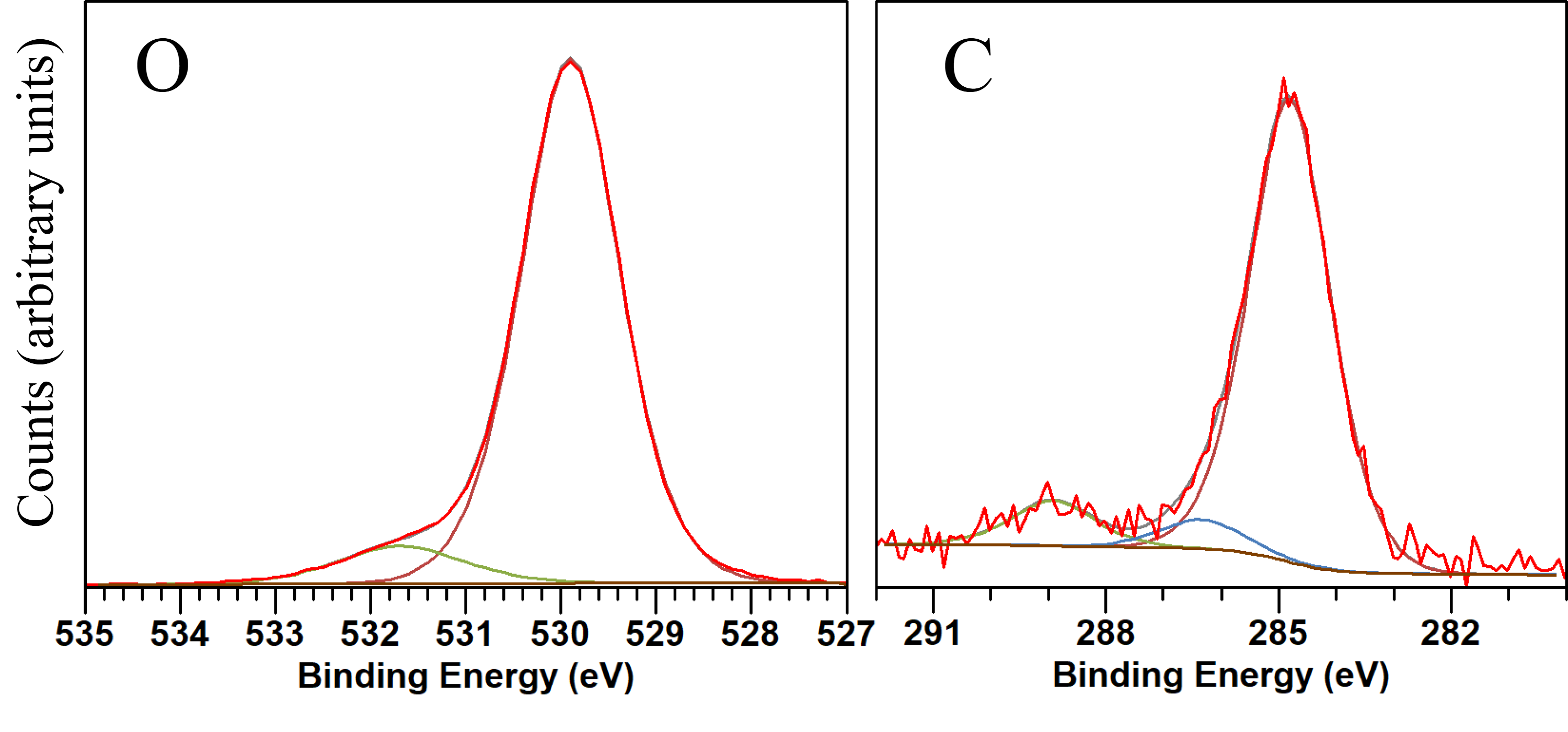}
\caption{C and O spectra (800C sample)}
\label{fig5:sub1}
\end{subfigure}%
\begin{subfigure}{.45\textwidth}
\centering
\includegraphics[width=0.9\textwidth]{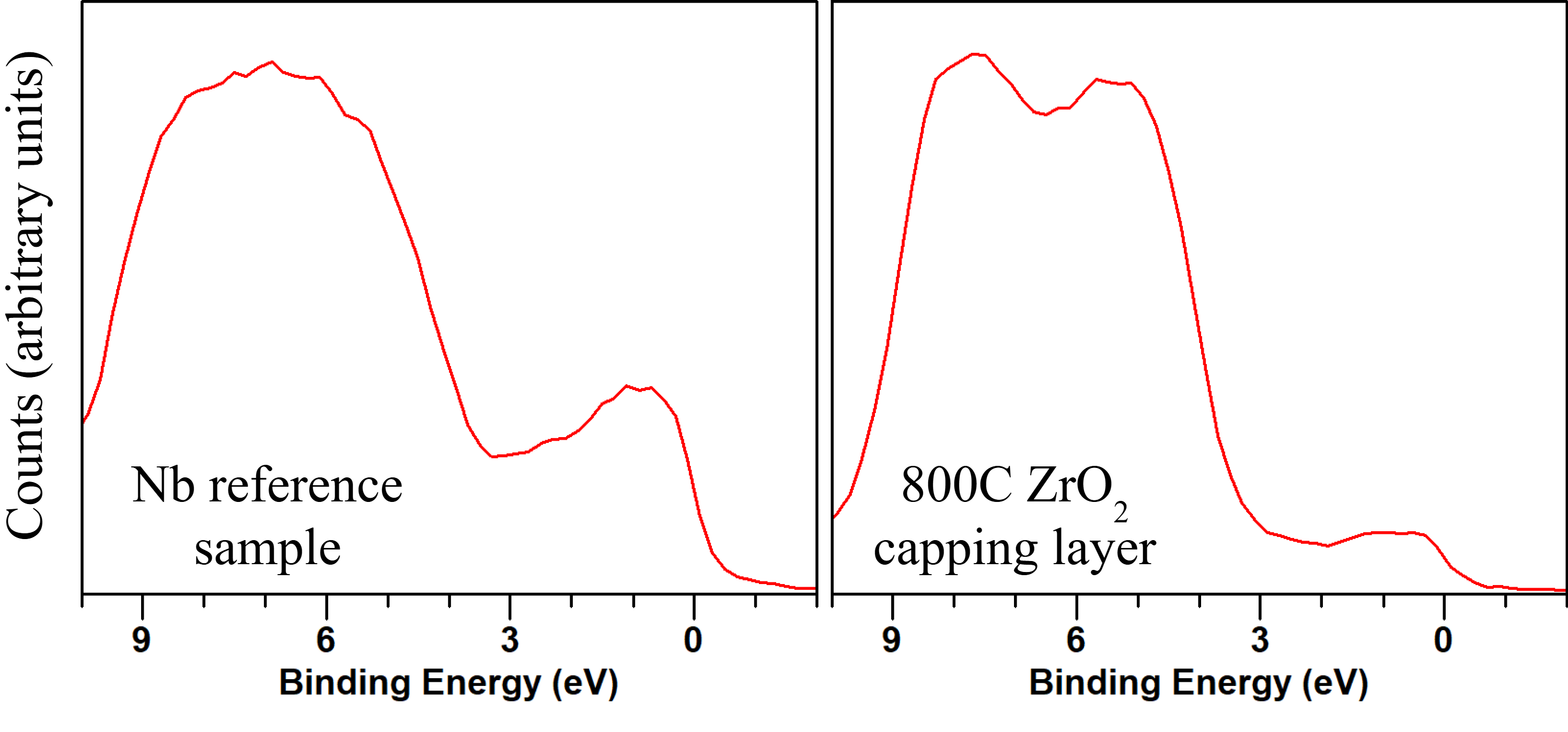}
\caption{Valence band spectra}
\label{fig5:sub2}
\end{subfigure}
\caption{C and O XPS spectra and peak deconvolution for the 800C-baked Zr capping layer, taken about 7 months after exposure to atmosphere (a). The O spectrum shows a large peak corresponding to ZrO$_2$, and a shoulder peak tentatively assigned to surface hydroxide groups and/or strongly adsorbed H$_2$O. The C spectrum shows adventitious carbon peaks, but no low-binding-energy peak that would indicate carbide formation. Valence XPS spectra for an electropolished Nb reference sample and for the 800C capping layer sample (b). For the valence spectra, the zero of binding energy has been set to the observed Fermi level, and large peaks correspond to valence band states of Nb$_2$O$_5$ and ZrO$_2$.}\label{fig5}
\end{figure}

In addition to Nb and Zr elemental spectra, we collected C and O elemental spectra, as well as valence band data for the 800C sample. We see no low-energy peak in the C spectrum, indicating an absence of metal carbides on the sample surface (Fig.~\ref{fig5:sub1}). This is notable because it is common for high-temperature-baked samples to have some carbide precipitates; fundamentally, these precipitates occur due to the relatively low solubility of carbon in niobium, which favors surface segregation and carbide formation at elevated temperatures~\cite{prudnikava2024situ}. In the oxygen spectrum we see a prominent peak corresponding to the oxygen atoms in the ZrO$_2$ layer, and a higher-energy shoulder peak which could indicate the presence of hydroxide groups and/or strongly-adsorbed water molecules on the sample surface~\cite{kamal2021core,kerber1996nature}. This would not come as a surprise, because ZrO$_2$ is a polar compound that can form strong hydrogen bonds with water molecules.

In the valence band data, we see large features significantly below the Fermi level for both the 800C sample and for an electropolished Nb reference sample; these features correspond to the ZrO$_2$ and Nb$_2$O$_5$ valence bands respectively (Fig.~\ref{fig5:sub2}). Near the Fermi level, we see features originating from metallic electrons in the substrate; these features are weaker in the 800C sample relative to the reference sample because the ZrO$_2$ capping layer is modestly thicker than the native Nb oxide.

In the Nb reference sample valence band spectrum, we see that the onset of the Nb$_2$O$_5$ valence band edge occurs about 4 eV below the apparent Fermi level. Because the expected bandgap of Nb$_2$O$_5$ is no more than 4 eV, this indicates that the Nb$_2$O$_5$ conduction band edge is very close to, if not below, the Fermi level. This is consistent with the presence of oxygen vacancies and associated conduction electrons in the Nb$_2$O$_5$ layer. The ZrO$_2$ valence band also occurs about 4 eV below the apparent Fermi level, but the ZrO$_2$ bandgap is generally expected to be 1-2eV larger than the Nb$_2$O$_5$ bandgap~\cite{portier2001relationships,dimitrov1996electronic}. This suggests that the ZrO$_2$ conduction band is significantly above the Fermi level in this sample, possibly indicating a lower defect density and fewer conduction-band electrons than the Nb$_2$O$_5$ native oxide. The ZrO$_2$ valence band also has a sharper edge than the Nb$_2$O$_5$ valence band, and has a double-peak structure characteristic of all common ZrO$_2$ phases~\cite{ZANDIEHNADEM198819,SORIANO1995659}. 

\begin{figure}[ht]
\centering
\begin{subfigure}{.45\textwidth}
\centering
\includegraphics[width=0.9\textwidth]{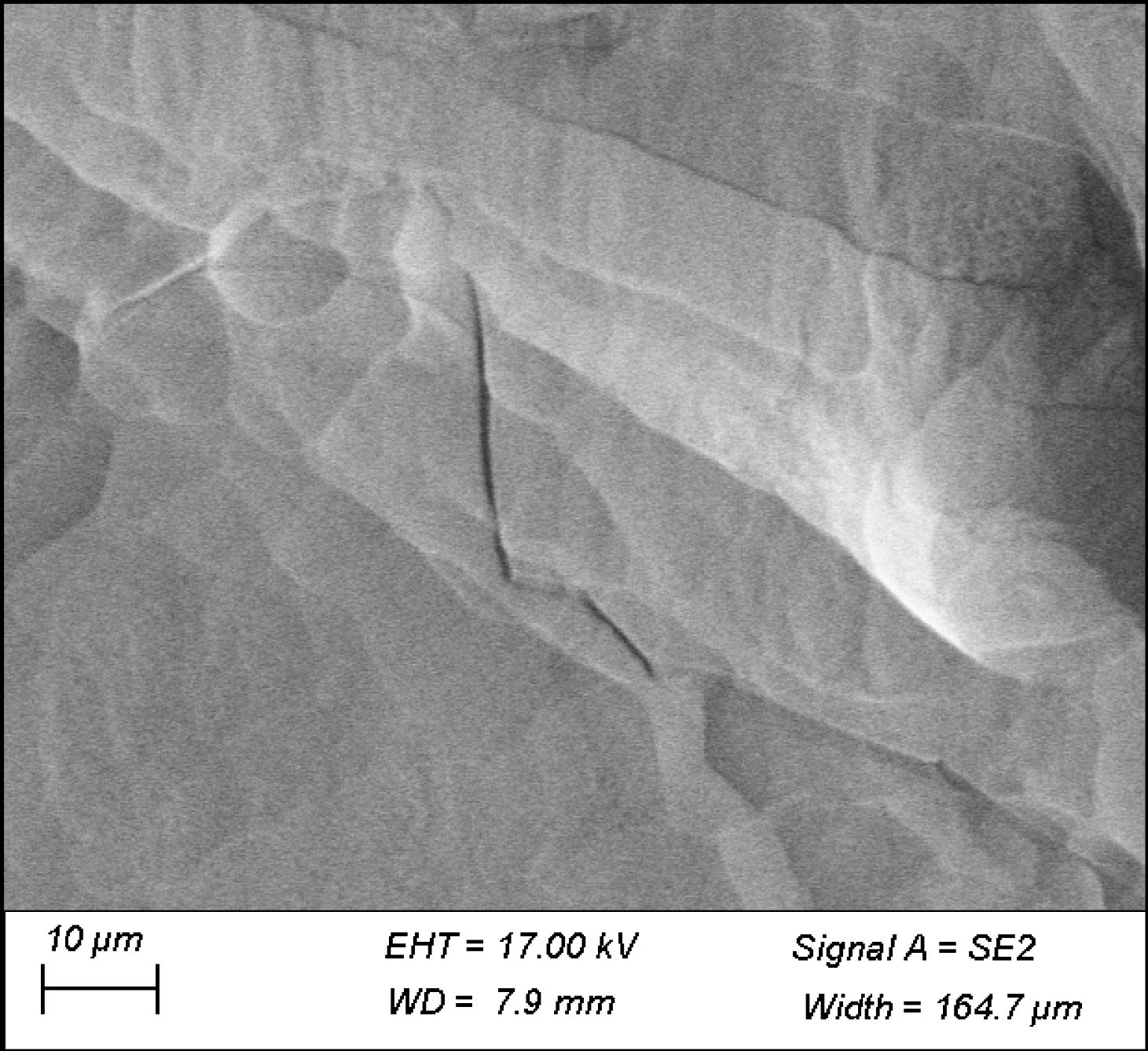}
\caption{SEM on $\mu$m lengthscale}
\label{fig6:sub1}
\end{subfigure}%
\begin{subfigure}{.45\textwidth}
\centering
\includegraphics[width=0.9\textwidth]{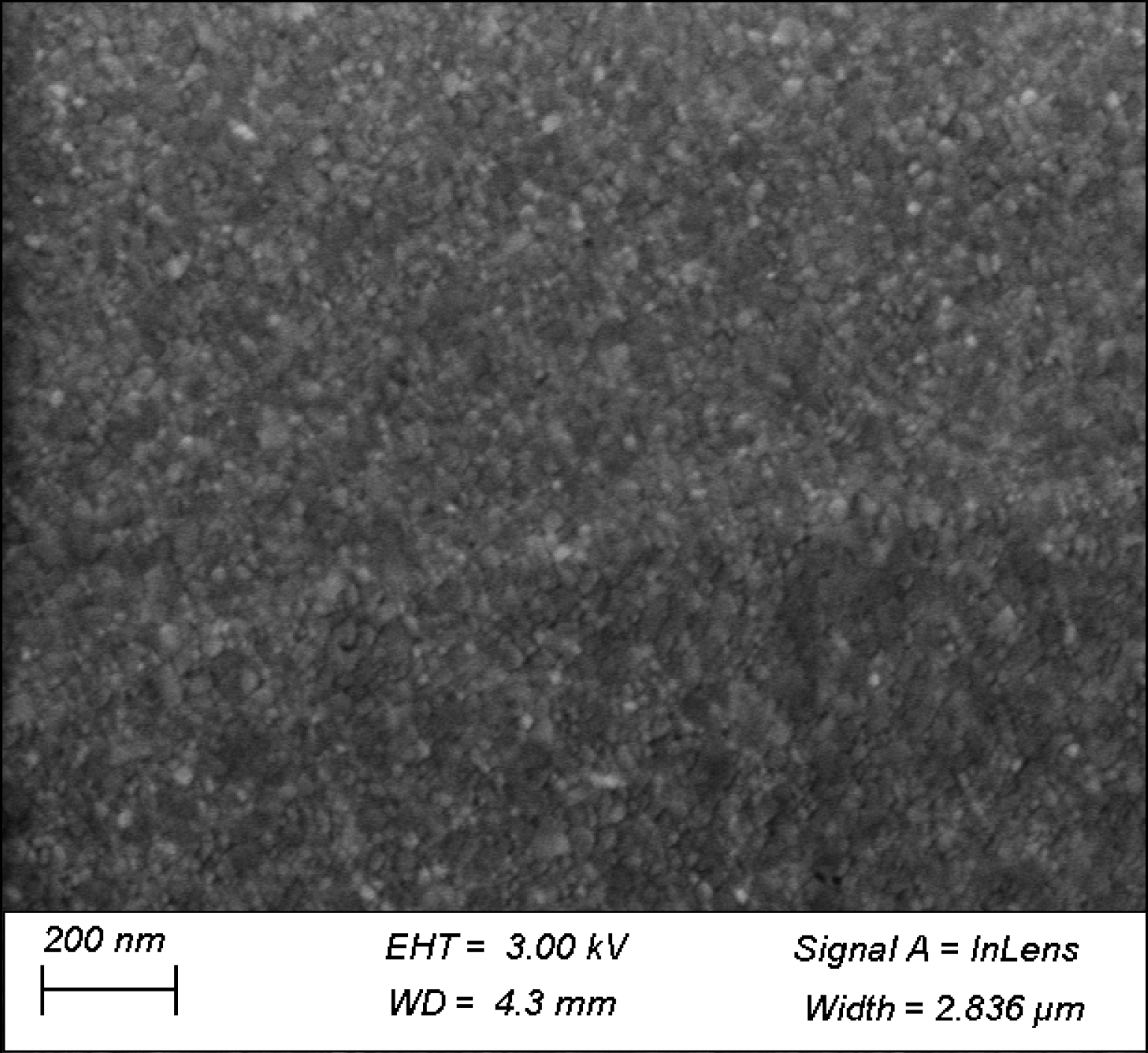}
\caption{SEM on nm lengthscale}
\label{fig6:sub2}
\end{subfigure}
\caption{Scanning electron microscopy (SEM) images of the 800C ZrO$_2$ capping layer sample, showing the grain structure of the Nb substrate on a micron lengthscale (a) and the grain structure of the capping layer on a nanometer lengthscale (b).}\label{fig6}
\end{figure}

\subsection{Scanning Electron Microscopy}\label{subsec3}

The 800C sample was further analyzed using scanning electron microscopy (SEM). On long lengthscales, the surface appears nearly featureless aside from substrate grain boundaries, consistent with the expected low-roughness of the electropolished Nb substrate (Fig.~\ref{fig6}). On shorter lengthscales, the SEM data shows a surface layer with well-defined crystal grains ranging from several nanometers to a few tens of nanometers in size. Based on XPS and STEM data, this crystalline layer is clearly identified as ZrO$_2$, and thus represents the first SEM evidence of crystallinity ever shown for an oxide on Nb. The layer appears to be fully continuous, with only slight modulations hinting at the presence of grain boundaries in the substrate.

\subsection{Cross-Section Scanning Transmission Electron Microscopy, Electron Energy Loss Spectroscopy, and Multislice Electron Ptychography}\label{subsec4}

\begin{figure}[ht]
\centering
\includegraphics[width=0.9\textwidth]{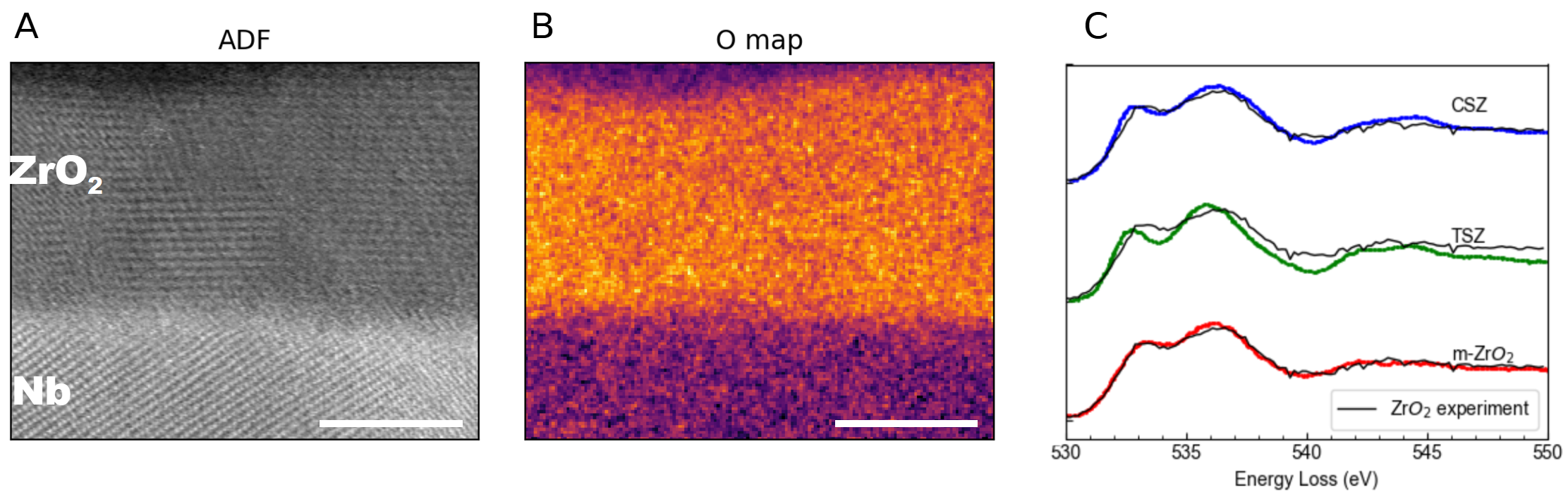}
\caption{STEM-EELS of the ZrO$_2$/Nb interface. Annular dark-field (ADF) image (A) and O K-edge EELS map (B) show the location of oxygen in the heterostructure. (C) O K-edge near edge fine structure of ZrO$_2$ (black) in comparison with cubic (blue), tetragonal (green), and monoclinic (red) phase references~\cite{mccomb1996bonding}. Scalebars: 5nm.}\label{fig7}
\end{figure}

A cross section taken from the 800C sample was analyzed using scanning transmission electron microscopy (STEM) and electron energy loss spectroscopy (EELS). The EELS data, focusing on the oxygen K-edge, clearly identifies the ZrO$_2$ capping layer in contrast to the oxygen-poor substrate (Fig.~\ref{fig7} A-B). Closer analysis of the shape of the EELS spectrum in comparison to data from the literature indicates that the capping layer at this location likely has a monoclinic crystal structure, as opposed to cubic or tetragonal~\cite{mccomb1996bonding}.

\begin{figure}[ht]
\centering
\includegraphics[width=0.8\textwidth]{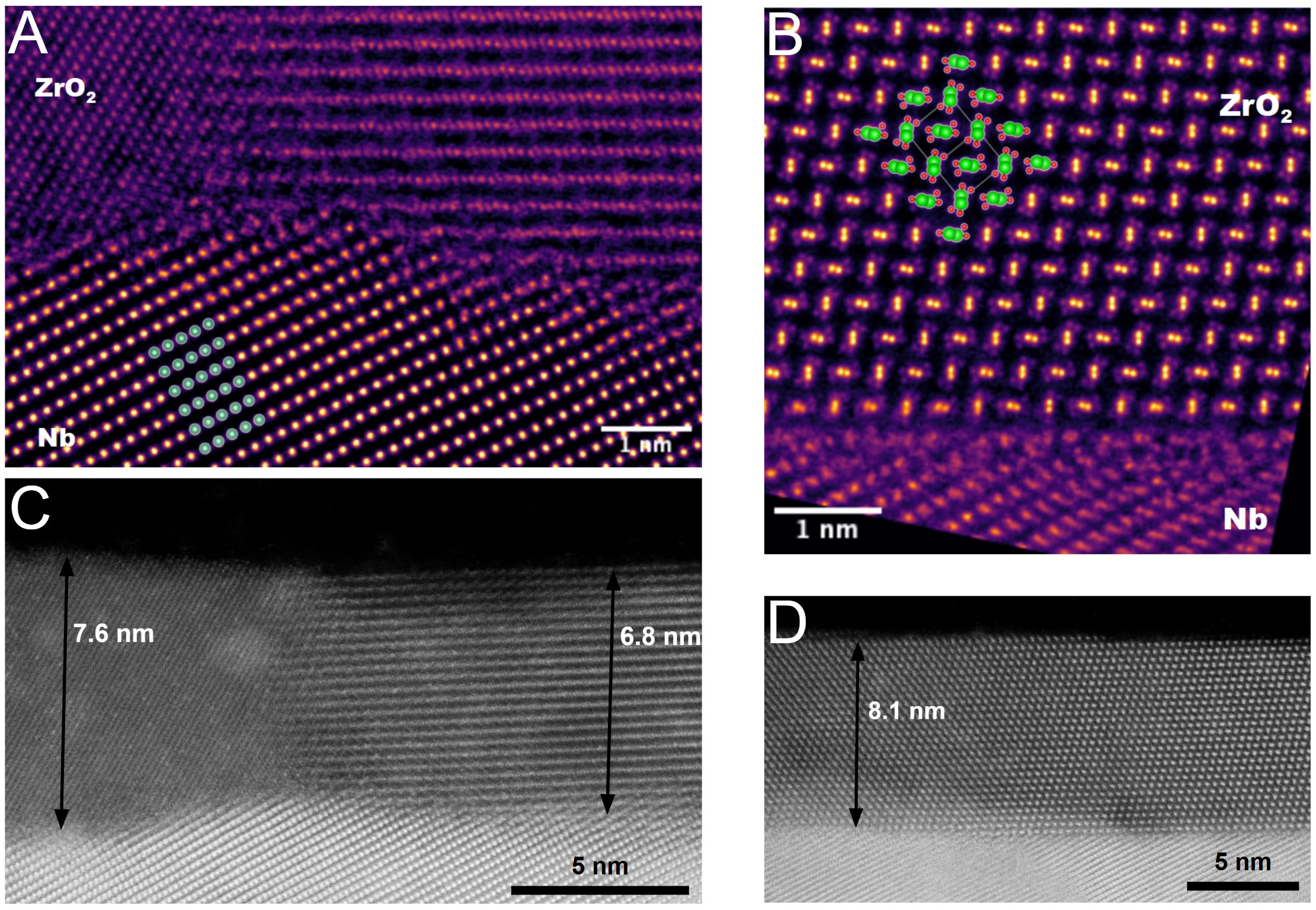}
\caption{Multislice electron ptychography (MEP) analysis of the interface beneath the 800C ZrO$_2$ capping layer in the vicinity of a ZrO$_2$ grain boundary (A), with the bcc Nb substrate oriented along its [110] zone axis. MEP analysis of the interface beneath a single uniform ZrO$_2$ grain (B), with the monoclinic ZrO$_2$ grain oriented along its [101] zone axis. ADF images C and D show the overall thickness of the capping layer in the regions surrounding A and B respectively.}\label{fig8}
\end{figure}

The ZrO$_2$ film appears polycrystalline with a lateral grain size in the range of 10 to 15 nm, roughly the same feature size visible over the larger region analyzed using SEM (Fig.~\ref{fig6}). Because of slight interface roughness, only a few locations along the Nb-ZrO$_2$ interface are sufficiently parallel to the direction of the electron beam to analyze them with sub-nm resolution. Two such regions of the interface were chosen for multislice electron ptychography (MEP) analysis, which reveals that the transition from ZrO$_2$ to Nb is much more abrupt than the transition from native-oxide Nb$_2$O$_5$ to Nb. In Fig.~\ref{fig8} A, the Nb substrate is along the [110] zone axis while the two ZrO$_2$ grains are off their high symmetry zone axis. In Fig.~\ref{fig8} B, a large ZrO$_2$ grain is aligned along its [101] zone axis. With both Zr and O atom columns resolved, the MEP image unambiguously confirms that this ZrO$_2$ grain is in the monoclinic phase, in agreement with EELS analysis in Fig.~\ref{fig7} C. Both regions in Fig.~\ref{fig8} A and B show sharp ZrO$_2$/Nb interfaces with 2 to 3 disordered atomic layers in between, which is about 2 to 5 \AA{} thick. The layer thickness is fairly uniform, ranging from approximately 7 to 8 nm as shown in the ADF images in Fig.~\ref{fig8} C and D.

While there does not appear to be an epitaxial relationship between the substrate and the capping layer, we nonetheless see that the bcc niobium crystal structure is in intimate contact with the monoclinic ZrO$_2$ crystal structure. The topmost atomic layer of the substrate shows some evidence of oxygen interstitials, while the bottommost layer of the capping layer shows some evidence of oxygen vacancies. This is consistent with XPS data, which showed very small but nonzero suboxide components for this sample. We conclude that a significant fraction of the Nb-ZrO$_2$ interface consists of an atomically-sharp transition from crystalline Nb to crystalline ZrO$_2$, the first observation of such an interface between a superconductor and a dielectric.

\section{Discussion}\label{sec3}

To understand the effectiveness of the ZrO$_2$ capping layer, we begin by considering the properties of the niobium native oxide. In the space of a few nanometers, the native oxide experiences an extraordinary gradient of oxygen chemical potential: from highly-oxidizing conditions on the surface (including not only exposure to atmosphere, but also for practical purposes exposure to humidity and/or high-pressure water rinsing), to highly-reducing conditions at the interface with niobium metal. In particular, at the elevated temperatures at which grain growth would occur in niobium oxide, high-purity niobium readily dissolves oxygen. This results in the dissolution of the niobium oxide layer into the substrate before it can fully crystallize ~\cite{prudnikava2024situ}.

For a candidate capping layer, we see that two key requirements can be used to screen for promising materials: stability at the capping layer surface, and stability at the interface with the niobium substrate. At the capping layer surface, the material must be stable in realistic atmospheric conditions: this rules out most highly-ionic oxides, which are generally attacked by water and water vapor due to their highly-polar surfaces~\cite{F29858100473,DIMITROV2002100,kuroda2000specific}. Additionally, the surface must be stable under vacuum annealing conditions necessary for capping layer crystallization. Even very slow evaporation rates on the order of single atomic layers per hour could be problematic, especially if they are grain-orientation dependent.

At the interface with the niobium substrate, the capping layer must retain oxygen under highly reducing conditions. The configurational entropy contribution to the free energy of oxygen dissolved in niobium grows linearly with temperature, making this requirement harder to satisfy at increasing temperatures. Our capping layer must satisfy this condition at a high enough temperature that crystallization can occur; this rules out oxides with formation energy too similar to that of the niobium oxide. Finally, we must consider that the capping layer could react with niobium to form intermetallic phases or ternary oxide phases; either of these events would likely result in unacceptable degradation of the capping layer and the formation of new defects.

\begin{figure}[ht]
\centering
\includegraphics[width=0.9\textwidth]{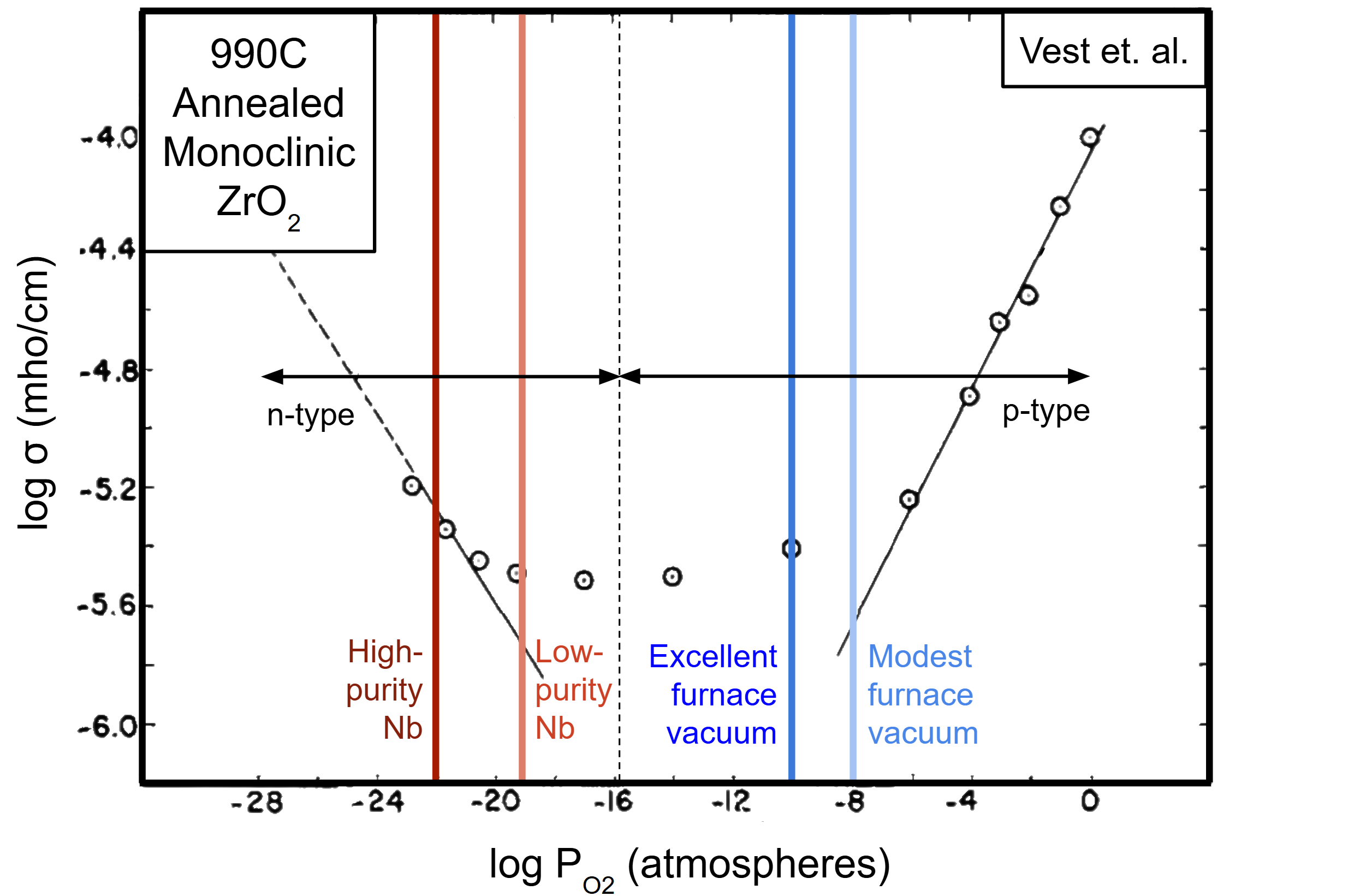}
\caption{ZrO$_2$ conductivity data adapted from Vest et. al. overlaid with oxygen chemical potential benchmarks relevant to vacuum furnace processing of capping layers on Nb~\cite{vest}. Based on this data, we expect our vacuum furnace process to produce a low-conductivity (low-defect) capping layer.}\label{fig9}
\end{figure}

Based on the microscopic analysis of our ZrO$_2$ layer, we see that this oxide is capable of meeting all necessary requirements for effective passivation of the niobium substrate. At the capping layer surface, while ZrO$_2$ is modestly more polar than the niobium native oxide, the amount of strongly-adsorbed water on the surface is evidently small based on our XPS results. More importantly, based on cross-section analysis, water does not appear to penetrate the bulk of the passivation layer or cause any other obvious form of degradation. Notably, other researchers who have studied ZrO$_2$ on Nb have subjected surfaces to high-pressure water rinsing without any obvious degradation~\cite{kalboussi2025crystallinity2}. Additionally, we find that the ZrO$_2$ layer does not evaporate at a detectable rate at the 800C temperature used to induce crystallization: our XPS analysis indicates that the 800C capping layer is precisely the same thickness as the 120C capping layer. This is in agreement with extrapolations of high-temperature ZrO$_2$ evaporation data from the literature, which suggest evaporation rates on the order of nanometers per millennia at 800C~\cite{hoch1954vapor}.

At the interface with the niobium substrate, our XPS analysis suggests very low concentrations of suboxide phases, and no indication of intermetallic phases. The cross-section data gives us an even clearer understanding of this interface, showing that the ZrO$_2$ capping layer indeed interfaces directly with the substrate, with no suboxide layer, ternary oxide phase, or intermetallic phase present. This is in agreement with experimental literature on the Nb-Zr system, which reports a low solubility for Zr in Nb at 800C and no low-energy intermetallic phases. Furthermore, the literature indicates that ZrO$_2$ has a formation energy per oxygen atom about 1.3 eV greater than that of the lowest Nb oxide, NbO~\cite{219851}. Given this energy difference and a process temperature of 800C, the dissolved oxygen concentration in the niobium substrate would need to be about a factor of $e^{1.3 eV/(k_B * T)} = 10^6$ below the oxygen solubility limit in Nb in order for the capping layer to dissolve; such low bulk oxygen concentrations do not occur even in ultra-high-purity samples.

Next, we consider the effects of annealing conditions on the defect structure of bulk ZrO$_2$. In general, the non-zero conductivity of ZrO$_2$ and other insulating oxides can be attributed to slight deficiencies in either oxygen (as observed in Nb$_2$O$_5$), leading to excess electrons in the conduction band and n-type semiconductor behavior, or in the metal species, leading to excess holes in the valence band and p-type semiconductor behavior. In addition to potentially adverse effects on RF performance, either type of defect could also hinder surface passivation by facilitating diffusion processes through the oxide.

Fortunately, the conductivity of ZrO$_2$ has been studied experimentally at temperatures and oxygen chemical potentials comparable to those involved in our 800C capping layer recipe~\cite{vest}. To estimate the oxygen chemical potential in the Nb substrate, we consider as a starting point the oxygen chemical potential of the lowest niobium oxide, NbO, which can be converted to units of $log(P_{O2})$ by means of a simple Boltzmann factor $e^{-E_{f}/(k_B{}*T)}$. Low-purity Nb (with oxygen content within an order of magnitude of the solubility limit) is estimated to have a slightly lower oxygen chemical potential than that of NbO. High-purity Nb comparable to what was used in our research has oxygen content about a factor of $10^3$ below that of low-purity Nb, corresponding to a difference of $-3$ in units of $log(P_{O2})$. Given these estimations, shown in Figure~\ref{fig9}, we expect low conductivity and thus low defect densities in ZrO$_2$.

Finally, we consider the question of the ZrO$_2$ crystal structure, which could have important implications for TLS defect concentrations and RF characteristics. Only monoclinic grains were found in the layer cross section, and no signs of defects were found in the regions of the cross section analyzed with MEP. However, the slight broadening of ZrO$_2$ XPS peaks suggests that some inhomogeneity may exist on a larger scale, either due to the presence of other ZrO$_2$ phases, or due to defects in the ZrO$_2$ layer that shift the conduction band edge relative to the Fermi level. Future research should also consider the possibility that the substrate grain texture (e.g. sputtered Nb vs. the larger-grain Nb used in this study) could affect ZrO$_2$ capping layer crystal structure, and thus could indirectly influence device performance.

\section{Conclusion}\label{sec5}

The development of a crystalline ZrO$_2$ capping layer on Nb with an atomically-sharp superconductor-substrate interface is a step toward minimizing interfacial losses in superconducting resonators. However, there is still much to understand about the properties of ZrO$_2$ capping layers, the structure of the Nb-ZrO$_2$ interface, and the interplay between material properties and low-field RF characteristics important for quantum applications. The thermodynamic stability of ZrO$_2$ allows for great freedom in exploring different conditions of vacuum annealing temperature and duration, which could influence grain size, crystal phase, and defect density, considering both defects at the interface between the substrate and the capping layer and precipitates at the sample surface. By fully exploring these possibilities, we can aim to achieve still-greater enhancements to device quality factors, and to deconvolve two-level system losses from intrinsic low-field, low-temperature loss mechanisms in quantum electronics.

\section{Methodology}\label{sec4}

\subsection{Sample Preparation}\label{subsec5}

Niobium samples were cut by waterjet from a sheet of 3-millimeter thick, high-RRR niobium sheet. After approximately 50 microns of chemical removal by buffered-chemical polishing and electropolishing, the samples were cleaned in an ultrasonic bath to remove any residual acid and air dried in a class-10 cleanroom to minimize exposure to adventitious carbon. The samples were then transported to the Cornell NanoScale Facility cleanroom for metallic zirconium deposition. An AJA-brand sputter deposition tool was used to first clean the sample surface with 300 seconds of argon sputtering, then (~20 seconds later, without breaking vacuum) deposit approximately 4 nanometers of metallic zirconium. Samples were then stored in a cleanroom atmosphere and annealed in a TM-brand high-vacuum furnace.

\subsection{X-Ray Photoelectron Spectroscopy}\label{subsec6}

X-Ray Photoelectron Spectroscopy (XPS) was performed using a Thermo Fisher Scientific Nexsa G2 XPS machine in the Cornell Center for Materials Research (CCMR). An electron flood gun was used to prevent surface charging, and a relatively low pass energy of 20 eV was used for elemental scans in order to effectively resolve small differences in binding energies between different oxidation states, while a pass energy of 50 eV was used for valence spectra.
Peak fitting was performed using CasaXPS software~\cite{fairley2021systematic}. Niobium metal peaks were fitted with a GL(75)T(1.5) peak shape, corresponding to a 25\% Gaussian-70\% Lorentzian blend with asymmetry specified by the T(1.5) parameter. Importantly, the higher-energy (3/2) doublet peak width was constrained to be 1.3 times the width of the lower-energy (5/2) doublet peak; this difference in doublet peak widths for metallic Nb is a result of the Super Coster Kronig effect~\cite{maartensson1981electron}. All Nb oxide peaks were fitted with a GL(50) peak shape, and all Nb peaks assumed a peak separation of 2.73 eV.
Zirconium metal peaks were fitted with a GL(70)T(1.2) peak shape, suboxide peaks were fitted with a GL(50) peak shape, and ZrO$_2$ peaks were fitted with a GL(40) peak shape. All Zr peaks assumed a peak separation of 2.385 eV, and additionally assumed an area ratio of 0.69 (instead of the usual $2/3$) for doublet peaks, following the work of~\cite{lackner2019using}.
All C and O peaks were fitted with a GL(50) peak shape.

\subsection{Scanning Electron Microscopy}\label{subsec8}

Scanning Electron Microscopy (SEM) was performed using a Zeiss Sigma 500 SEM in the Cornell Center for Materials Research. Images were taken using the in-lens secondary electron detector.

\subsection{Cross-Section Transmission Electron Microscopy and Electron Energy Loss Spectroscopy}\label{subsec9}

The TEM lamella used in this study was prepared with a Thermo Fisher Helios Focused ion beam using the standard lift-out method. The milling of the lamella was done with voltages starting from 30 kV down to 2 kV during the final milling step. The HAADF–STEM images, 4D-STEM datasets for MEP and EELS data were acquired with an aberration-corrected Thermo Fisher Spectra 300 at 300 keV beam energy and 30 mrad convergence semi-angle. 
4D-STEM datasets for MEP were collected EMPAD-G2 detector ~\cite{philipp2022very}. The scan was performed with a step size of 0.4 Å over a 256 × 256 raster, with a maximum collection semi-angle of 51 mrad. The probe was focused approximately 10 nm above the sample surface. Reconstruction was performed in PtyRAD ~\cite{lee2025ptyrad} using four mixed probe modes and 15 layers of 1 nm thickness each. Each reconstruction was iterated for over 5000 iterations.

The EELS data were acquired with ThermoFisher Selectris-X camera. We used an electron current of 48 pA, and EELS collection angle of 33 mrad. The energy dispersion was chosen to be 0.17 eV. The field of view is 13.4 × 12.8 nm$^2$ with about 0.4 Å scanning step size. The spectra were not deconvolved, and the average thickness in the field of view is 0.23 inelastic mean free path.

\section{Author Contributions}\label{sec6}

N. Sitaraman conceptualized the development of ZrO$_2$ capping layers, fabricated samples, performed XPS experiments, and drafted the manuscript. Z. Baraissov performed EELS and STEM experiments. A. Grassl performed SEM experiments. Z.Baraissov, D. Tong and H.Yang performed the ptychographic reconstructions. M. Liepe and D. Muller supervised and guided the research. All authors contributed to editing and revision of the manuscript.

\section{Data Availability}\label{sec7}

The data that support the findings of this study are available from the corresponding author upon reasonable request.

\section{Acknowledgements}\label{sec8}

This work was supported by the US National Science Foundation under Award PHY-1549132, the Center for Bright Beams and the US Department of Energy under Award DE-SC0024907. The authors thank Dr. Michael Schmid for advice on the analysis of Zr XPS spectra.

\bibliography{bibliography}

\end{document}